\documentclass[12pt]{article}
\usepackage{booktabs}
\usepackage{float}
\usepackage{amssymb,amsmath,amsfonts, mathtools}

\usepackage{enumerate}
\usepackage{dsfont}
\usepackage[sort,compress]{cite}
\usepackage{verbatim}
\usepackage{amsfonts,euscript,amssymb,stmaryrd,braket}
\usepackage{graphics,tikz}
\usetikzlibrary{arrows,decorations.markings,patterns}
\usepackage{caption}
\usepackage{subcaption}
\usepackage{slashed}
\definecolor{hyperref}{RGB}{026,028,185}
\usepackage[bookmarks=true,colorlinks=true,linkcolor=hyperref,citecolor=hyperref,urlcolor=hyperref,bookmarksnumbered]{hyperref}
 \usepackage{booktabs}
 \usepackage{multirow}
\usepackage{tabularx}
\usepackage{longtable}
\usepackage{bbold,bbding}
\usepackage{amsthm} 
\usepackage{graphicx}

 \def\clock{{\count0=\time
           \divide\count0 60
           \ifnum\count0<10 0\fi\the\count0
           \multiply\count0 -60 \advance\count0 \time
           :\ifnum\count0<10 0\fi \the\count0
         }}
\newcommand{\timestamp}{{\small\vbox{\hbox{\tt\jobname.tex}
\hbox{\the\day/\the\month/\the\year, \clock}}}}

\newcommand{\be}{\begin{equation}}
\newcommand{\ee}{\end{equation}}
\newcommand{\ba}{\begin{eqnarray}}
\newcommand{\ea}{\end{eqnarray}}
\newcommand{\no}{\nonumber}

\newcommand{\IndA}{M}
\newcommand{\IndB}{N}
\newcommand{\IndC}{P}
\newcommand{\IndD}{Q}
\newcommand{\IndE}{R}
\newcommand{\IndF}{S}
\newcommand{\ms}{1}

\def\e{\epsilon}

\def\rmp{{\rm p}}

\newcommand{\arsinh}{\operatorname{arsinh}}

\begin{document}
\begin{center}

\begin{flushright} 
\footnotesize
HU-EP-17/32
\end{flushright}%

\bigskip
\bigskip

{\Large \bf On regularizing the AdS superstring worldsheet }

\bigskip

{\large Valentina Forini}

\medskip

\small 

{           Institut f\"ur Physik, Humboldt-Universit\"at zu Berlin\\
            IRIS Adlershof, Zum Gro\ss en Windkanal 6,\\
            D-12489 Berlin, Germany\\
             valentina.forini@physik.hu-berlin.de\\
}
\bigskip

\end{center}

\begin{abstract}
In this short review (to appear as a contribution to an edited volume)
we discuss perturbative and non-perturbative approaches to the quantization of the Green-Schwarz string in AdS backgrounds with RR-fluxes,  where the guiding thread is the
use of genuine field theory methods, the search for a good regularization scheme associated to them  
and  the generality of the analysis carried out. We touch upon various computational setups, both analytical and numerical, and on the role of their outcomes in understanding the detailed structure of the AdS/CFT correspondence.

\end{abstract}

\tableofcontents


\section{Introduction }
\renewcommand{\theequation}{1.\arabic{equation}}
\setcounter{equation}{0}

Over the previous decade there has been beautiful progress in obtaining exact results in the framework of the duality between superconformal gauge theories and string theory in $AdS$ backgrounds with Ramond-Ramond (RR) fluxes, or AdS/CFT correspondence. 
 Several examples of physical observables exist by now, whose functional behavior with the coupling 
is known  -- explicitly or implicitly -- not only in the regimes which are naturally under control perturbatively (both from a gauge theory and sigma-model perspective) but also at finite coupling. Essentially two methods are decisive here, 
the first relying on the integrability of the underlying system~\cite{Beisert:2010jr} 
 and the second on supersymmetric localization~\cite{Pestun:2007rz,Pestun:2016zxk}.  

However,  not only integrability is in the finite coupling region  an assumption, and supersymmetric localization is only accessible in a limited set of cases (for those observables protected by supersymmetry~\footnote{See however~\cite{Correa:2012at} for a relevant extension of this set.}). Importantly, from the point of view of the string worldsheet theory -- which is ours in this note -- integrability is a solid fact only classically, and supersymmetric localization is not even formulated. 
The Green-Schwarz superstring on AdS backgrounds with RR fluxes  remains, beyond its supergravity approximation, a complicated interacting two-dimensional field theory which presents subtleties also at perturbative level.
%
Its action, when explicitly expanded in terms of independent fermionic degrees of freedom, is highly non-linear and usually quantized in a semiclassical approach~\cite{Gubser:2002tv,Frolov:2002av}, expanding around a classical solution in powers of the (effective) string tension~\cite{McLoughlin:2010jw}. Here difficulties may arise due to the fact that fermionic string coordinates, which are spacetime spinors,  appear in the Lagrangian always through their two-dimensional projection involving derivatives of the classical background (which, in order to define fermion propagators and perform perturbation theory, must be non-trivial). Such derivatives introduce a dimensional scale and appear nonlinearly in the quartic fermionic terms,  leading to non-renormalizable interactions and higher-power divergences beyond one-loop~\footnote{This is already true for the flat space case~\cite{Polyakov:2004br}. See also discussion in~\cite{Roiban:2007dq}.}.  Verifying the cancellation of the UV divergences with suitable regularization schemes -- crucial for a well-defined expansion -- may be then non-trivial. 
The search of regularization which is ``good'', \emph{i.e.} equivalent to the one (implicitly) assumed by the calculations performed via integrability or localization, characterizes the work  reviewed in the first part of this note. 
Quantizing the theory in a semi-classical approximation implies, beyond the leading order which defines minimal string surfaces (to be suitably regularized at the $AdS$ boundary), solving the spectral problem of highly non-trivial differential operators of Laplace and Dirac type, namely evaluating the zeta-function determinant in the context of elliptic boundary value problems, a procedure which below we illustrate on two relevant examples.  We also sketch the evaluation of next-to-leading corrections, or two loop order, and comment on an efficient alternative to Feynman diagrammatics, based on unitarity cuts,  which may be used in the case of on-shell objects such as worldsheet scattering amplitudes. 

 

At a non-perturbative level, a natural way to regularize a theory and perform ab initio calculations within it is to define it on a discretized spacetime or lattice. Lattice field theory methods have been recently become a subject of study also in the framework of worldsheet string models~\cite{McKeown:2013vpa,Forini:2016sot,Bianchi:2016cyv}.   This approach bypasses the subtleties of realizing supersymmetry on the lattice -   which characterise the lattice approach to the duality from the gauge theory side~\cite{Catterall:2014vga} - in that the Green-Schwarz superstring formulation that we use displays supersymmetry only in the target space. In the two-dimensional string world-sheet model under analysis  supersymmetry appears as a flavour symmetry.  Importantly, local symmetries (diffeomorphism and fermionic kappa-symmetry) are all fixed, and only scalar fields   (some of which anti-commuting) appear, assigned to sites. This rather simplified setting -  useful to have at most quartic fermionic interactions - still retains the sophisticated dynamics of relevant observables in this framework. 


Below we will be mostly dealing with the $AdS_5\times S^5$ superstring;  with few exceptions, the majority of the observations generalizes to other AdS/CFT relevant backgrounds.

\section{Sigma-model and perturbation theory\label{sec:perturbative}} 

\renewcommand{\theequation}{2.\arabic{equation}}
\setcounter{equation}{0}

When evaluating  the $AdS_5\times S^5$ string partition function in a semiclassical quantization,  it is possible and useful to remain extremely general at least in writing down the fluctuation spectrum about such solution, applying elementary concepts of intrinsic and extrinsic geometry to the properties of string worldsheet embedded in a $D$-dimensional curved space-time. Taking full advantage of the equations of Gauss, Codazzi, and Ricci for  surfaces embedded in a general background one obtains simple and general expressions for perturbations over them~\footnote{This follows and enlarges earlier investigations~\cite{Callan:1989nz,Drukker:2000ep}.}. For example, writing down the complete mass matrix $\mathcal{M}$ in the bosonic fluctuation sector only requires as an input generic properties of the classical configuration,  basic information about the space-time background and the inclusion of a suitable choice of orthonormal vectors which are orthogonal to the surface spanned by the string solution 
\be
\label{Mtot}
\begin{split}
\mathcal{M}_{ij} =&  -m^2_{AdS_5}  (\hat N_i\cdot \hat N_j)-m^2_{S^5}\,(\bar N_i \cdot\bar N_j)+K_{i\alpha\beta} K_{j}^{\alpha\beta},\\
m^2_{AdS_5}\equiv&  \gamma^{\rho\sigma} ( \hat t_{\rho}\cdot\hat t_{\sigma} )\  \ \ \ \ \  \mathrm{and}\ \ \ \ \ \ 
m^2_{S_5}\equiv -\gamma^{\rho\sigma} (\bar t_{\rho} \cdot \bar t_{\sigma} ).
\end{split}
\ee
Above, $\gamma_{\alpha\beta}\,,\,\alpha,\beta=1,2$ is the induced metric (pullback of the $AdS_5\times S^5$ target space metric), hats and bars refer to the projections onto $AdS_5$ and $S^5$ of vectors tangent ($t$) and orthogonal ($N$) to the worldsheet,
 $K^i_{\alpha\beta}\equiv K^A_{\alpha\beta} N^{i}_{A}$ is the extrinsic curvature of the embedding, $i,j\dots,8$ are transverse space-time indices.  
 
To proceed in the one-loop analysis, one is to explicitly compute the functional determinants associated to the fluctuations operators. This is in general difficult, 
except in the case of rational rigid string solutions, so-called ``homogeneous''~\cite{Frolov:2003qc,Frolov:2003tu,Frolov:2004bh,Park:2005ji,Beisert:2005mq}, for which the Lagrangian has coefficients constant in the worldsheet coordinates and the one-loop partition function results  in a sum over characteristic frequencies which are relatively simple to calculate. 
For the non-homogenuous case, one is to restrict to problems which are effectively one-dimensional. 
This step, which may involve  regularization subtleties and other  issues -- as the appropriate definition of integration measure, kappa-symmetry ghosts, Jacobians due to change of fluctuation basis --
is often feasible with standard techniques, such as  the Gelfand-Yaglom method for the evaluation of functional determinants (stated originally in \cite{Gelfand:1959nq} and later improved in~\cite{Forman1987, Forman1992, McKane:1995vp, Kirsten:2003py, Kirsten:2004qv, Kirsten:2007ev})~\footnote{See for example~\cite{Kirsten:2004qv,Dunne:2007rt}, or the concise review in Appendix B of~\cite{Forini:2015bgo}.}. This algorithm has the advantage of computing ratios of determinants bypassing the computation of the full set of eigenvalues and is based on the solution of an auxiliary initial value problem.
Considering the pair of $n$-order ordinary differential operators in one variable
\be 
\mathcal{O}=P_0(\sigma) \frac{d^n}{d\sigma^n}+\sum_{k=0}^{n-1} P_{n-k}(\sigma) \frac{d^k}{d\sigma^k}\,,
~~
\hat{\mathcal{O}}=P_0(\sigma) \frac{d^n}{d\sigma^n}+\sum_{k=0}^{n-1} \hat
{P}_{n-k}(\sigma) \frac{d^k}{d\sigma^k}
\ee
with coefficients being $r\times r $ complex matrices, continuous functions of $\sigma$ on the finite interval $I=[a,b]$. The principal symbol  (proportional to the coefficient $P_0(\sigma)$ of the highest-order derivative) is assumed to be the same for both operators, and invertible ($\textrm{det}P_0(\sigma)\neq0$) on the whole interval~\footnote{This assumption ensures that the leading behaviour of the eigenvalues is comparable, thus the ratio is well-defined despite the fact each determinant is formally the product of infinitely-many eigenvalues of increasing magnitude. }. 
The operators act on the space of square-integrable $r$-component functions $\bar{f}\equiv\left(f_1, f_2, ..., f_r\right)^T\in\mathcal{L}^2\left(I\right)$, 
and $nr\times nr$ constant matrices $M,N$ implement the linear boundary conditions at the extrema of $I$
\be 
\label{bc_GY}
M \left(\begin{array}{cc}
\bar{f}\left(a\right)\\
\frac{d}{d\sigma}\bar{f}\left(a\right)\\
\vdots\\
\frac{d^{n-1}}{d\sigma^{n-1}}\bar{f}\left(a\right)
\end{array}\right)+
N \left(\begin{array}{cc}
\bar{f}\left(b\right)\\
\frac{d}{d\sigma}\bar{f}\left(b\right)\\
\vdots\\
\frac{d^{n-1}}{d\sigma^{n-1}}\bar{f}\left(b\right)
\end{array}\right)=
\left(\begin{array}{cc}
0\\
0\\
\vdots\\
0
\end{array}\right).
\ee
The particular significance of the Gel'fand-Yaglom theorem is that it drastically reduces the complexity of finding the spectrum of the operators of interests
\begin{gather}
\mathcal{O} \bar{f}_{\lambda}(\sigma)=\lambda \bar{f}_{\lambda}(\sigma)\,,
\quad\qquad
\hat{\mathcal{O}} \hat{\bar{f}}_{\hat{\lambda}}(\sigma)=\hat{\lambda} \hat{\bar{f}}_{\hat{\lambda}}(\sigma),
\end{gather}
encoding it into the elegant formula for (e.g. even-order) differential operators
 \begin{gather}
\label{ratio_forman_even}
\frac{\textrm{Det}_\omega \mathcal{O}}{\textrm{Det}_\omega \hat{\mathcal{O}}}=
\frac{\exp\left\{\frac{1}{2} \int_a^b \textrm{tr} \left[P_1(\sigma) P_0^{-1}(\sigma)\right] d\sigma\right\} \textrm{det}\left[M+N Y_{\mathcal{O}}\left(b\right)\right]}
{\exp\left\{\frac{1}{2} \int_a^b \textrm{tr} \left[\hat{P}_1(\sigma) P_0^{-1}(\sigma)\right] d\sigma\right\} \textrm{det}\left[M+N Y_{\hat{\mathcal{O}}}\left(b\right)\right]}\,.
\end{gather}
This result  agrees with the  one obtained via $\zeta-$function regularization for elliptic differential operators. Above,  the $nr\times nr$ matrix
\begin{gather}
\label{GY_Y}
Y_{\mathcal{O}}(\sigma)=\left(\begin{array}{cccc}
\bar{f}_{(I)}(\sigma) & \bar{f}_{(II)}(\sigma) & \dots & \bar{f}_{(nr)}(\sigma)\\
\frac{d}{d\sigma}\bar{f}_{(I)}(\sigma) & \frac{d}{d\sigma}\bar{f}_{(II)}(\sigma) & \dots & \frac{d}{d\sigma}\bar{f}_{(nr)}(\sigma)\\
\vdots & \vdots & \ddots & \vdots \\
\frac{d^{n-1}}{d^{n-1}\sigma}\bar{f}_{(I)}(\sigma) & \frac{d^{n-1}}{d^{n-1}\sigma}\bar{f}_{(II)}(\sigma) & \dots & \frac{d^{n-1}}{d^{n-1}\sigma}\bar{f}_{(nr)}(\sigma)\\
\end{array}\right)
\end{gather}
uses all the independent homogeneous solutions of
\begin{gather}
\mathcal{O}\bar{f}_{(i)}(\sigma)=0 \qquad i=I, II, ..., 2r
\end{gather}
chosen such that $Y_{\mathcal{O}}\left(a\right)=\mathbb{I}_{nr}$.  
In a number of relevant cases this method has been strikingly efficient in combination with the underlying classical integrable structure of the Green-Schwarz superstring on $AdS_5\times S^5$, revealed by the presence of a class of integrable  differential operators -- tipically of Lam\'e type~\cite{Beccaria:2010ry, Forini:2014kza} -- for which solutions are known in the literature. 
%
%
In some other problems highly non-trivial second-order \emph{matrix} 2d differential operators appear, whose coefficients have a complicated coordinate-dependence, for example in the (effectively bosonic) mixed-modes case of a  folded string spinning   in $S^5$
with two large angular momenta $(J_1,J_2)$, solution of the Landau-Lifshitz (LL)  effective action 
of \cite{Kruczenski:2003gt}. In this case one has to build the ingredients of the Gelf'and Yaglom method, studying ex novo fourth order differential equations with doubly periodic coefficients~\cite{Forini:2014kza}. Among other findings, this study allows the
\emph{analytic} proof of equivalence between the full  exact one-loop string partition function (for the one-spin folded string) in conformal and static gauge  -- a non-trivial statement which finds its counterpart only in flat space~\cite{Fradkin:1982ge}.

\bigskip


The computation of the disc partition function for the $AdS_5\times S^5$ superstring appears to be subtle, beyond the supergravity approximation, in the cases of classical solutions corresponding to supersymmetric Wilson loops.
For euclidean minimal surfaces  ending at the boundary on circular loops -- the maximal 1/2 BPS~\cite{Drukker:2000ep, Kruczenski:2008zk,Buchbinder:2014nia}, the 1/4  BPS family of ``latitudes''~\cite{Drukker:2005cu,Drukker:2006ga,Drukker:2007qr,Forini:2015mca,Forini:2015bgo,Faraggi:2016ekd}, the k-wound case in the fundamental representation~\cite{Bergamin:2015vxa}, as well as loops in 
k-symmetric and k-antisymmetric representations~\cite{Buchbinder:2014nia} -- the first correction to the partition function gives a result which disagrees with the gauge theory result, conjectured in~\cite{Erickson:2000af, Semenoff:2002kk, Drukker:2000rr,Drukker:2005kx} and proven in~\cite{Pestun:2007rz,Pestun:2009nn} via supersymmetric localization. 
To eliminate ambiguities due to the absolute normalization of the string partition function, and under the assumption that the latter is independent on the geometry of the classical worldsheet,
one should consider the ratio between the partition functions for two supersymmetric Wilson loops with the same topology. 
This was done in~\cite{Forini:2015bgo}~\footnote{See also~\cite{Faraggi:2016ekd}.}, where the one-loop determinants for fluctuations about the classical solutions corresponding to a generic ``latitude'' - the 1/4 BPS Wilson loops of~\cite{Drukker:2005cu,Drukker:2006ga,Drukker:2007qr} - and the maximal 1/2-BPS circle were evaluated with the Gel'fand-Yaglom method, where a disagreement with the exact gauge theory result was confirmed. 
Recent developments suggest that to reconcile sigma-model perturbation theory and localization one may consider such ratio and use heat kernel techniques in a perturbative approach  about the case of the maximal circle~\cite{Heat_kernel_perturbative}\footnote{A recent study has clarified how the agreement is reached also accounting for an  IR anomaly associated with the divergence in the conformal factor, carefully defining an invariant cutoff and proceeding in the evaluation of functional determinants directly evaluting phaseshifts for all the fluctuation modes~\cite{Cagnazzo:2017sny}.}. The relevant string worldsheet for the maximal circle is $AdS_2$, where heat kernel explicit expressions for the spectra of Laplace and Dirac operators are available~\cite{Camporesi:1990wm, Camporesi:1994ga,Camporesi:1992tm,Camporesi:1995fb}, 
and in this case one explicitly evaluates their  corrections due to the near-$AdS_2$ geometry induced by the  generic latitude in $S^2\subset  S^5$ parametrized by a small angle $\theta_0$.  
Then  one considers the perturbative expansion 
\begin{equation}\label{expansion}
\begin{split}
g_{ij}&=\bar{g}_{ij}+\theta_0^2~\tilde{g}_{ij}+O\left(\theta_0^4\right)\\
\mathcal{O}&= \bar{\mathcal{O}}+\theta_0^2\,\tilde{\mathcal{O}}+O\left(\theta_0^4\right)\,,\\
K_{\mathcal{O}}(x,x^{'};t) & = \bar{K}_{\mathcal{O}}(x,x^{'};t)+\theta_0^2~\tilde{K}_{\mathcal{O}}(x,x^{'};t)+O\left(\theta_0^4\right)
\end{split}
\end{equation}
in the heat equation
\begin{eqnarray}\label{heatequation}
 \left(\partial_{t}+\mathcal{O}_{x}\right)K_{\mathcal{O}}(x,x^{'};t)=0\qquad\quad K_{\mathcal{O}}(x,x^{'};0)=\frac{1}{\sqrt{g}}\delta^{\left(d\right)}\left(x-x^{'}\right)\mathbb{I}~,
\end{eqnarray}
and finds for the correction $\tilde{K}_{\mathcal{O}}$ to the functional trace $K_{\mathcal{O}}\left(t\right) =\bar{K}_{\mathcal{O}}\left(t\right)+\theta_0^2\,\tilde{K}_{\mathcal{O}}\left(t\right)+O\left(\theta_0^4\right)$
\be\label{heat_trace_perturb}
\tilde{K}_{\mathcal{O}}\left(t\right)  =-t\int_{x}\sqrt{\bar{g}}~~\textrm{tr}\left[\tilde{\mathcal{O}}_{x}~\bar{K}_{\mathcal{O}}\left(x,x^{'};t\right)\right]_{x=x^{'}}\,.
\ee
This translates in the perturbative evaluation of each determinant in the partition function as~\cite{Heat_kernel_perturbative} 
\begin{eqnarray}\label{det_perturb}
\log(\det\mathcal{O}) &=& -\bar\zeta_{\mathcal{O}}^{'}\left(0\right)-\theta_0^2\,\tilde\zeta_{\mathcal{O}}^{'}\left(0\right)+O(\theta_0^4)\,,\\\nonumber\\ \nonumber
\bar{\zeta}_{\mathcal{O}}\left(s\right)  &=&\frac{1}{\Gamma\left(s\right)}\int_{0}^{\infty}dt\,t^{s-1}\,{\bar K}_{\mathcal{O}}\left(t\right)\,,~~~
\tilde{\zeta}_{\mathcal{O}}\left(s\right)  =\frac{1}{\Gamma\left(s\right)}\int_{0}^{\infty}dt\,t^{s-1}\tilde{K}_{\mathcal{O}}\left(t\right)~.
\end{eqnarray}
This approach turns our to be successful: the gauge theory exact result  is indeed reproduced, at one loop and at order $\mathcal{O}(\theta_0^2)$,  by the analysis in sigma-model perturbation theory. ~
Despite being both based on zeta-function regularization, the two procedures illustrated here for the evaluation of functional determinants differ substantially on few aspects. 
In this context, where the spectral problem is effectively (after Fourier-transforming in, say,  $\tau$) one-dimensional, the  Gelfand-Yaglom uses a zeta-function-like regularization in $\sigma$ -- whose outcome is equivalent to the solution~\eqref{ratio_forman_even} above -- and a cutoff- regularization in the sum over the Fourier $\tau$-modes, and is therefore not a diffeo-invariant regularization scheme. It also requires considering ratios of determinants for differential operators with the same principal symbol, which in turns implies  a functional rescaling by the conformal factor, and the introduction of a fictitious boundary -- a  cut at the origin of the disk --  introduced in~\cite{Kruczenski:2008zk,Forini:2015bgo, Faraggi:2016ekd} to allow the calculation of determinants on the finite interval (see also~\cite{Frolov:2004bh, Dekel:2013kwa}). Such regulator does not appear in the heat kernel approach, which a fully two-dimensional method~\footnote{We refer to the very recent~\cite{Cagnazzo:2017sny} for the explicit account of the role played by conformal rescalings and invariant cutoff regulators in solving the disagreement with the gauge theory result.}.

\subsection{Higher orders}

\begin{figure}[h]
    \centering
               \includegraphics[scale=0.5]{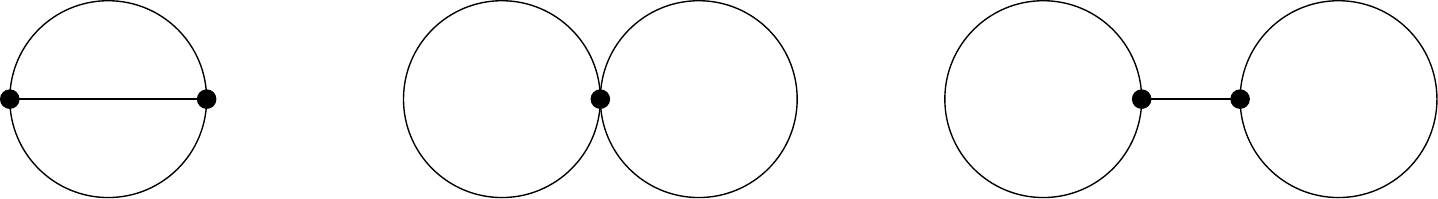}
                \caption{Sunset, double bubble and double tadpole appearing in the two-loop contribution to the string partition function for the cusped light-like solution~\cite{Giombi:2009gd,Bianchi:2014ada}. } 
             \label{diagrams}
\end{figure}

Beyond one-loop, one has to further restrict the class of feasible  problems to homogenous configurations, and trade the standard conformal gauge with the so-called $AdS$  lightcone gauge, where the light-cone is entirely in $AdS$~\cite{Metsaev:2000yu}. 
This setup -- where propagators are in general simple, and (in the bosonic case)  diagonal, a fact that drastically reduces the number of Feynman diagrams to be evaluated. -- was efficiently used in~\cite{Giombi:2009gd} to evaluate the strong coupling corrections to the $\mathcal{N}=4$ SYM cusp anomaly up to two-loop order. In~\cite{Bianchi:2014ada} a very similar calculation was done in the considerably more involved case of the $AdS$  lightcone gauge-fixed action derived via double dimensional reduction from a 
$D=11$ membrane action  based on the supercoset $OSp(8|4)/(SO(7)\times SO(1, 3))$. As the relevant classical solution is homogeneous, the one-loop partition function is a sum of simple frequencies. At two  loops, the possible topologies of connected vacuum diagrams (sunset, double bubble, double tadpole)  occurring when studying the effective string action at two loops in this setup are in Fig.~\ref{diagrams}.  
When combining vertices and propagators in the sunset diagrams  various non-covariant integrals are originated, but standard reduction techniques allow to rewrite every integral as a linear combination of the two following scalar ones
\begin{align}\label{integrals}
I\left(m^2\right) & \equiv \int \frac{d^2p}{\left(2\pi\right)^2}\, \frac{1}{p^2+m^2} \\
I\left(m_1^2,m_2^2,m_3^2\right) & \equiv \int \frac{d^2p\, d^2q\,d^2r}{\left(2\pi\right)^4}\,  \frac{\delta^{(2)}(p+q+r)}{(p^2+m_1^2)(q^2+m_2^2)(r^2+m_3^2)} \quad. 
\end{align}
In this process it is standard~\cite{Giombi:2009gd,Roiban:2007jf,Roiban:2007dq,Bianchi:2014ada} to set to zero power UV divergent massless tadpoles, as in dimensional regularization 
\begin{equation}\label{dimreg}
\int \frac{d^2p}{(2\pi)^2}\, \left( p^2\right)^n = 0\,, \qquad\qquad n\geq 0\,,
\end{equation}
so that all manipulations in the numerators  are performed in $d=2$, which has the advantage of simpler tensor integral reductions. 
While UV finiteness is not obvious, as each diagram in \eqref{integrals} is separately divergent (the last one in the IR, the former in both UV and IR), all  logarithmically divergent integrals -- remaining after the power-like are set to zero via \eqref{dimreg} -- happen to cancel out in the computation and there is no need to pick up an explicit regularization scheme to compute them.
Such non-trivial result, together with establishing the quantum consistency of the string action  proposed in~\cite{Uvarov:2009hf, Uvarov:2009nk}, has been the first non-trivial check at strong coupling of the conjectured~\cite{Gromov:2014eha} 
 all-order  expression  of the interpolating function $h(\lambda)$ appearing in terms of which all calculations based on the integrability of the AdS$_4$/CFT$_3$ system are based. 

\subsection{Unitarity methods in $d=2$ dimensions}

\begin{figure}[h]
    \centering
               \includegraphics[scale=0.5]{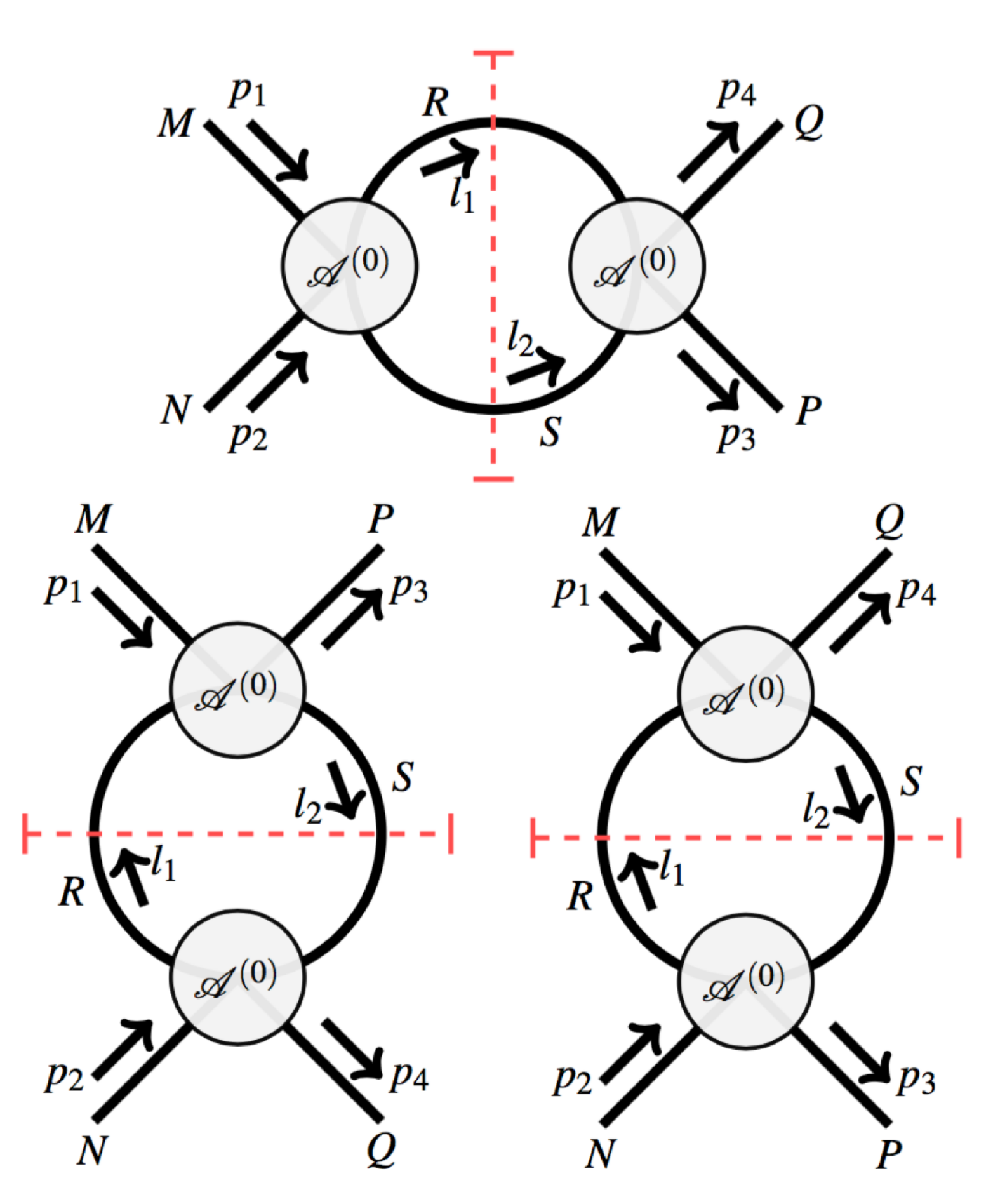}
                \caption{Diagrams representing s-, t- and u-channel cuts contributing to the four-point one-loop amplitude. } 
             \label{stu}
\end{figure}

As extremely  efficient alternative to Feynman diagrammatics -- however only well-stablished for on-shell objects -- unitarity-cut techniques are a powerful tool in non-abelian gauge theories for the evaluation of space-time scattering amplitudes (see e.g.~\cite{Roiban:2010kk}).
In~\cite{Bianchi:2013nra,Engelund:2013fja} their use was initiated for the one-loop, perturbative study of the S-matrix for  massive two-dimensional field theories, describing the scattering of the Lagrangian excitations.
Here the method boils down to a reverse application of Cutkowsky rules, allowing the extraction of  the discontinuity of a Feynman diagram across its branch cut. 
In applying the standard unitarity rules (derived from the optical theorem)~\cite{Bern:1994zx} 
to the example of a one-loop four point amplitude, one considers \emph{two-particle cuts}, obtained by putting two intermediate lines on-shell. The contributions that follow to the imaginary part of the amplitude are therefore given by the sum of $s$- $t$- and $u$-  channel cuts illustrated in Fig.~\ref{stu},  explicitly 
\begin{eqnarray}\no
\!\!\!\!\!
\mathcal{A}^{(1)}{}^{\IndC\IndD}_{\IndA\IndB}(p_1,p_2,p_3,p_4)|_{s-cut}=\int\frac{d^2 l_1}{(2\pi)^2}\int\frac{d^2 l_2}{(2\pi)^2}\ i\pi\delta^+({l_1}^2-\ms)\ i\pi\delta^+(l_2^2-\ms)\\\no
\!\!\!\!\!\!\!\!\!\!\times\,\mathcal{A}^{(0)}{}_{\IndA\IndB}^{\IndE\IndF}({p_1,p_2,l_1,l_2})\mathcal{A}^{(0)}{}_{\IndF\IndE}^{\IndC\IndD} ({l_2,l_1,p_3,p_4})
\\\no
\!\!\!\!\!
\mathcal{A}^{(1)}{}^{\IndC\IndD}_{\IndA\IndB}(p_1,p_2,p_3,p_4)|_{t-cut}=\int\frac{d^2 l_1}{(2\pi)^2}\int\frac{d^2 l_2}{(2\pi)^2}\ i\pi\delta^+({l_1}^2-\ms)\ i\pi\delta^+({l_2}^2-\ms)\\\no
\!\!\!\!\!\!\!\!\!\!\times\,\mathcal{A}^{(0)}{}_{\IndA\IndE}^{\IndF\IndC}({p_1,l_1,l_2,p_3})\mathcal{A}^{(0)}{}_{\IndF\IndB}^{\IndE\IndD}({l_2,p_2,l_1,p_4})\\\no
\!\!\!\!\!
\mathcal{A}^{(1)}{}^{\IndC\IndD}_{\IndA\IndB}(p_1,p_2,p_3,p_4)|_{u-cut}=\int\frac{d^2 l_1}{(2\pi)^2}\int\frac{d^2 l_2}{(2\pi)^2}\ i\pi\delta^+({l_1}^2-\ms)\ i\pi\delta^+({l_2}^2-\ms)\\\no
\!\!\!\!\!\!\!\!\!\!\times\,\mathcal{A}^{(0)}{}_{\IndA\IndE}^{\IndF\IndD}({p_1,l_1,l_2,p_4})\mathcal{A}^{(0)}{}_{\IndF\IndB}^{\IndE\IndC}({l_2,p_2,l_1,p_3})
\end{eqnarray}
where $\mathcal{A}^{(0)}$ are tree-level amplitudes and a sum over the complete set of intermediate states $\IndE,\IndF$ (all allowed particles for the cut lines) is understood.  
%


Notice that tadpole graphs,  having no physical two-particle cuts,  are by definition ignored in this procedure. 
To proceed, in each case one uses  the  momentum conservation at the vertex involving the momentum $p_1$ to integrate over $l_2$, e.g. for the $s$-channel
\ba
 &&\!\!\!\!\!\!\!\!\!\!\!\!\!\!\!\!\!\!\!\!\!\!\!
 \mathcal{\widetilde{A}}^{(1)}{}^{\IndC\IndD}_{\IndA\IndB}(p_1,p_2,p_3,p_4)|_{s-cut}=\int\frac{d^2 l_1}{(2\pi)^2}\,i\pi\delta^+({l_1}^2-\ms)\, i\pi\delta^+(({l_1}-{p_1}-{p_2})^2-\ms)\nonumber\\\no
&&\qquad\times \,\,\widetilde{\mathcal{A}}^{(0)}{}_{\IndA\IndB}^{\IndE\IndF}({p_1,p_2,l_1,-l_1+p_1+p_2}) \,\widetilde{\mathcal{A}}^{(0)}{}_{\IndF\IndE}^{\IndC\IndD}({-l_1+p_1+p_2,l_1,p_3,p_4})\,.\label{2_8} 
\ea
The simplicity of the two-dimensional kinematics and of being at one loop plays now its role, since in each of the  integrals the set of zeroes of the $\delta$-functions is a discrete set, and the cut loop-momenta are frozen to specific values~\footnote{ At two loops, to constrain completely the four components of the two momenta circulating in the loops  one needs four cuts, each one giving an on-shell $\delta$-function. Two-particle cuts at two loops would result in a manifold of conditions for the loop momenta.}. This allows us to pull out the tree-level amplitudes with the loop-momenta evaluated at those zeroes~\footnote{This is like using $f(x)\delta(x-x_0)=f(x_0)\delta(x-x_0)$, where $f(x)$ are the tree-level amplitudes in the integrals.}. In what remains, following standard unitarity computations~\cite{Bern:1994zx}, we apply the replacement $i\pi \delta^+(l^2-1) \longrightarrow \tfrac{1}{l^2-1}$ (\emph{i.e.} the Cutkowsky rule in reverse order) which sets loop momenta back off-shell, thus reconstructing scalar bubbles.  
 This allows  to rebuild, from its imaginary part, the cut-constructible piece of the amplitude and of the S-matrix, via~\cite{Bianchi:2013nra}
\be\label{AandS}
S_{\IndA\IndB}^{\IndC\IndD}(p_1,p_2)\equiv \frac{J(p_1,p_2)}{4\e_1 \e_2} \widetilde{\mathcal{A}}_{\IndA\IndB}^{\IndC\IndD}(p_1,p_2,p_1,p_2) ~.
\ee
where the Jacobian $J(p_1,p_2)=1/(\partial \e_{\rmp_1}/\partial \rmp_1-\partial\e_{\rmp_2}/\partial \rmp_2)$ depends on the dispersion relation $\e_\rmp$, on-shell energy associated to $\rmp$ (the \emph{spatial} momentum) for the theory at hand. 
The expression for the 
one-loop S-matrix elements is given by the following simple sum of products of two tree-level amplitudes~\footnote{In \eqref{eqn:final}, $\tilde S^{(0)}(p_1,p_2)=4 (\e_2\,\rmp_1-\e_1\,\rmp_2) S^{(0)}(p_1,p_2)$ and the denominator on the right-hand side comes from the Jacobian $J(p_1,p_2)$ assuming a standard relativistic dispersion relation (for the theories we consider, at one-loop this is indeed the case). } 
\begin{eqnarray} \label{eqn:final}
&& {S^{(1)}}{}_{\IndA\IndB}^{\IndC\IndD}(p_1,p_2)=\frac{1}{4 (\e_2\,\rmp_1-\e_1\,\rmp_2)}\,\Big[ {\tilde S^{(0)}}{}_{\IndA\IndB}^{\IndE\IndF}(p_1,p_2){\tilde S^{(0)}}{}_{\IndE\IndF}^{\IndC\IndD}(p_1,p_2)\,I_{p_1+p_2}\\ \nonumber\vphantom{\frac{1}{4 (\e_2\,\rmp_1-\e_1\,\rmp_2)}}
 &&+ {\tilde S^{(0)}}{}_{\IndA\IndE}^{\IndF\IndC}(p_1,p_1){\tilde S^{(0)}}{}_{\IndF\IndB}^{\IndE\IndD}(p_1,p_2)\,I_0 
 \vphantom{\frac{1}{4 (\e_2\,\rmp_1-\e_1\,\rmp_2)}}
 +{\tilde S^{(0)}}{}_{\IndA\IndE}^{\IndF\IndD}(p_1,p_2){\tilde S^{(0)}}{}_{\IndF\IndB}^{\IndC\IndE}(p_1,p_2)\,I_{p_1-p_2}\,\Big]
\end{eqnarray}
where the coefficients are given in terms of the bubble integral
\be
I_p=\int \frac{d^2 q}{(2\pi)^2} \frac{1}{(q^2-\ms+i\e) ((q-p)^2-\ms+i\e)}~
\ee 
 and read explicitly~\footnote{The $t$-channel cut requires a prescription~\cite{Bianchi:2013nra}.} 
\begin{eqnarray}\nonumber
 I_{p_1+p_2}=\frac{i\pi-\arsinh(\e_2\,\rmp_1-\e_1\,\rmp_2)}{4\pi i\,(\e_2\,\rmp_1-\e_1\,\rmp_2)},~~
 I_0=\frac{1}{4\pi i},~~
 I_{p_1-p_2}=\frac{\arsinh(\e_2\,\rmp_1-\e_1\,\rmp_2)}{4\pi i\,(\e_2\,\rmp_1-\e_1\,\rmp_2)}~. 
\end{eqnarray}
As it only involves the scalar bubble integral in two dimensions,  the result~\eqref{eqn:final} following from our procedure is inherently \emph{finite}. No additional regularization is required and the result can be compared directly with the $2 \to 2$ particle S-matrix (following from the finite or renormalized four-point amplitude) found using standard perturbation theory.
Of course, this need not be the case for the original bubble integrals before cutting -- due to factors of loop-momentum in the numerators. These divergences, along with those coming from tadpole graphs, which are not considered in this procedure, should be taken into account for the renormalization of the theory. We have not investigated this issue, since all the theories considered in~\cite{Bianchi:2013nra}~\footnote{They include, among relativistic theories, a class of generalized sine-Gordon models, defined by a gauged WZW model for a coset G/H plus a potential. Notable non-relativistic cases are the superstring worldsheet models in $AdS_5\times S^5$  and $AdS_3 \times S^3 \times M^4$~\cite{Bianchi:2014rfa}.}   are either UV-finite or renormalizable. 
The method, applied to various models, has shown enough evidence to postulate that supersymmetric, integrable two-dimensional theories should be cut-constructible via standard unitarity methods. For bosonic theories with integrability,  agreement was found with perturbation theory up to a finite shift in the coupling. In the case of the superstring worldsheet models in $AdS_5\times S^5$~\cite{Bianchi:2013nra} and $AdS_3 \times S^3 \times M^4$~\cite{Bianchi:2014rfa}, the method  allowed non-trivial confirmations of the integrability prediction together with conjectures (then confirmed) on the one-loop phases. 

\renewcommand{\theequation}{3.\arabic{equation}}
\setcounter{equation}{0}

\section{$AdS_5\times S^5$ superstring on a lattice}
\label{sec:lattice}
\renewcommand{\theequation}{4.\arabic{equation}}
\setcounter{equation}{0}

The natural, genuinely field-theoretical way to investigate the finite-coupling region and in general the non-perturbative realm of a quantum field theory  is to discretize the spacetime where the model lives, and proceed with numerical methods for the  lattice field theory so defined. A rich and interesting program of putting $\mathcal{N}=4$ SYM a space-time lattice is being carried out for some years~\cite{Kaplan:2005ta,Catterall:2014vka,Joseph:2015xwa,Schaich:2015ppr,Bergner:2016sbv}~\footnote{See also the numerical, non-lattice formulation of  $\mathcal{N}=4$ SYM on  $R\times S^3$ as plane-wave (BMN) matrix model~\cite{Ishii:2008ib, Ishiki:2008te,Ishiki:2009sg, Hanada:2010kt,Honda:2011qk,Honda:2013nfa,Hanada:2013rga}.}. 
Alternatively, one could discretize the worldsheet spanned by the Green-Schwarz string embedded in $AdS_5\times S^5$.   
 If the aim is a test of the AdS/CFT correspondence and/or the integrability of the string sigma model, it is is obviously computationally cheaper to use a two-dimensional grid, rather than a four-dimensional one, where no gauge degrees of freedom are present and all fields are assigned to sites - so that only scalar fields (some of which anticommuting) appear in the relevant action.  Also, although we are dealing with superstrings, there is here no subtlety involved with putting supersymmetry on the lattice (see e.g.~\cite{Catterall:2014vga}),  both because of the Green-Schwarz formulation of the action (with supersymmetry only manifest in the target space) 
and because $\kappa$-symmetry is gauge-fixed.  In general, one merit of this analysis  is to explore another route  via which  lattice simulations~\footnote{See for example~\cite{Hanada:2016jok} and reference therein on possible further uses of lattice techniques in  AdS/CFT.} could become a potentially  efficient tool in numerical holography.  
Following the earlier proposal of~\cite{McKeown:2013vpa}, such a route has been taken in~\cite{Forini:2016sot, Bianchi:2016cyv} to investigate relevant observables in AdS/CFT,  discretizing the dual two-dimensional string worldsheet. There, the focus is on  particularly important observables completely ``solved'' via integrability~\cite{Beisert:2006ez}: the cusp anomalous dimension of $\mathcal{N}=4$ SYM -- measured by the path integral of an open string bounded by a null-cusped Wilson loop at the AdS boundary -- and the spectrum of excitations around the corresponding string minimal surface. The relevant string worldsheet theory, an AdS-lightcone gauge-fixed action~\cite{Metsaev:2000yf,Metsaev:2000yu},  is a highly non-trivial 2d non-linear sigma-model with rich non-perturbative dynamics. On the lattice, several subtleties appear (fermion doublers, complex phases) which require special treatment, as we sketch below.

\subsection{The observable in the continuum}

The cusp anomaly of $\mathcal{N}=4$ SYM governs the renormalization of a cusped Wilson loop, and according to AdS/CFT
should be represented by the  path integral of an open string ending on the loop at the AdS boundary 
\be\label{Z_cusp}
\begin{split}
 \langle W[C_{\rm cusp}]\rangle\equiv  Z_{\rm cusp}&= \int [D\delta X] [D\delta\Psi]\, e^{- S_{\rm cusp}[X_{\rm cl}+\delta X,\delta\Psi]} \\
&= e^{-\Gamma_{\rm eff}}\equiv e^{-\frac{1}{8} f(g)\,V_2 }~.                         
\end{split}
\ee
Above, $X_{\rm cl}=X_{\rm cl}(t,s)$ - with $t,s$ the temporal and spatial coordinate spanning the string worldsheet -  is the relevant classical solution~\cite{Giombi:2009gd}, $S_{\rm cusp}[X+\delta X,\delta\Psi]$ is the action for field fluctuations over it -- the fields being both bosonic and fermionic string coordinates $X(t,s),~\Psi(t,s)$ -- and is reported below in equation \eqref{S_cusp} in terms of the effective bosonic and fermionic degrees of freedom remaining after gauge-fixing. Being an homogenous solution, the worldsheet volume  simply factorizes out~\footnote{The normalization of $V_2$ with a $1/4$ factor follows the convention of~\cite{Giombi:2009gd}.} in front of the function of the coupling  $f(g)$, as in the last equivalence above.  
%
Rather than partition functions, in a lattice approach it is natural to study  vacuum expectation values. In simulating  the vacuum expectation value of the ``cusp'' action  
\begin{eqnarray}\label{vevaction}
\langle S_{\rm cusp}\rangle&=& \frac{ \int [D\delta X] [D\delta\Psi]\, S_{\rm cusp}\,e^{- S_{\rm cusp}}}{ \int [D\delta X] [D\delta\Psi]\, e^{- S_{\rm cusp}}} = -g\,\frac{d\ln Z_{\rm cusp}}{dg}\equiv g\,\frac{V_2}{8}\,f'(g)   ~,
  \end{eqnarray}
one therefore obtains information on the \emph{derivative} of the scaling function. 
In the continuum, the   $AdS_5\times S^5$   superstring   action $S_{\rm cusp}$ describing quantum  fluctuations around  the null-cusp background is~\cite{Giombi:2009gd} (after Wick-rotation) 
\begin{eqnarray}\nonumber
&& \!\!\!\!\!\!\!\!\!\!\!\!
S_{\rm cusp}=g \int dt ds~ \textstyle\Big\{ 
|\partial_{t}x+\textstyle{\frac{1}{2}}x|^{2}+\frac{1}{ {z}^{4}} |\partial_{s} {x}-\textstyle{\frac{1}{2}} {x}|^{2}-\textstyle{\frac{1}{{z}^{2}}}\left( {\eta}^{i}{\eta}_{i}\right)^{2}+\frac{1}{ {z}^{4}}\left(\partial_{s} {z}^{M}-\textstyle{\frac{1}{2}} {z}^{M}\right)^{2} \\\nonumber
 && \!\!\!\!\!\!\!\!\!\!\!\!
\textstyle 
+\textstyle{\big(\partial_{t}z^{M}+\frac{1}{2} {z}^{M}+\frac{i}{ {z}^{2}} {z}_{N} {\eta}_{i}\left(\rho^{MN}\right)_{\phantom{i}j}^{i} {\eta}^{j}\big)^{2}}
  +i\left( {\theta}^{i}\partial_{t}{\theta}_{i}+ {\eta}^{i}\partial_{t}{\eta}_{i}+ {\theta}_{i}\partial_{t}{\theta}^{i}+ {\eta}_{i}\partial_{t} {\eta}^{i}\right) \\\nonumber
 &&  \!\!\!\!\!\!\!\!\!\!\!\!
 \textstyle 
 +2i\Big[\textstyle{\frac{1}{z^{3}}}z^{M} {\eta}^{i}\left(\rho^{M}\right)_{ij}
 \left(\partial_{s} \theta^j-\textstyle{\frac{1}{2}} \theta^j-\frac{i}{{z}} {\eta}^{j}\left(\partial_{s} {x}-\frac{1}{2} {x}\right)\right)\\\label{S_cusp}
&& \textstyle{+\frac{1}{{z}^{3}}{z}^{M}{\eta}_{i} (\rho_{M}^{\dagger} )^{ij}\left(\partial_{s}{\theta}_{j}-\frac{1}{2}{\theta}_{j}+\frac{i}{{z}}{\eta}_{j}\left(\partial_{s}{x}-\frac{1}{2}{x}\right)^{*}\right)\Big]\,\Big\}}
\end{eqnarray}
Above, $x,x^*$ are the two bosonic $AdS_5$ (coordinate) fields transverse  to the $AdS_3$ subspace of the classical solution, and $z^M\, (M=1,\cdots, 6)$ are the bosonic coordinates of the $AdS_5\times S^5$ background in Poincar\'e parametrization, with  $z=\sqrt{z_M z^M}$, remaining after fixing the AdS light-cone gauge. 
 The fields $\theta_i,\eta_i,\, i=1,2,3,4$ are  4+4 complex anticommuting variables for which  $\theta^i = (\theta_i)^\dagger,$ $\eta^i = (\eta_i)^\dagger$. They transform in the fundamental representation of the $SU(4)$ R-symmetry and do not carry (Lorentz) spinor indices.  The matrices $\rho^{M}_{ij} $ are the off-diagonal
blocks of $SO(6)$ Dirac matrices $\gamma^M$ in the chiral representation
and
$(\rho^{MN})_i^{\hphantom{i} j} = (\rho^{[M} \rho^{\dagger N]})_i^{\hphantom{i} j}$ are  the
$SO(6)$ generators. 
In \eqref{S_cusp} -- where a massive parameter $m\sim P^+$, usually set to one, is restored -- local bosonic (diffeomorphism) and fermionic ($\kappa$-) symmetries originally present have been fixed. With this action one can directly proceed to the perturbative evaluation of the cusp anomaly  ($K$ is the Catalan constant) and of the dispersion relation for the field excitations 
\begin{equation}\label{cuspperturbative}
\begin{split}
f(g)&=4\,g\,\Big(1-\frac{3\log2}{4\pi\,g}-\frac{K}{16\,\pi^2\,g^2}+\mathcal{O}(g^{-3})\Big) \\
m^2_{ x}(g)&=\frac{m^2}{2}\,\Big(1-\frac{1}{8 \,g}+\mathcal{O}(g^{-2})\Big)~.
\end{split}
\end{equation}
%
%
%
While the bosonic part of~\eqref{S_cusp} can be easily discretized and simulated, Gra\ss mann-odd fields are 
formally integrated out, letting their determinant to become part -- via exponentiation in terms of pseudo-fermions, see \eqref{fermionsintegration} below -- of the Boltzmann weight of each configuration in the statistical ensemble. In the case of higher-order fermionic interactions -- as in \eqref{S_cusp}, where they are at most quartic -- this is possible via the introduction of auxiliary fields realizing a linearization. The most natural linearization~\cite{McKeown:2013vpa} introduces  $7$ real auxiliary  fields, one scalar $\phi$ and a  $SO(6)$ vector field $\phi_M$, with a  Hubbard-Stratonovich transformation~
\begin{eqnarray}\label{HubbardStratonovich}
&& \!\!\!\!\!\!\!
e^{ -g\int dt ds  \Big[-\textstyle{\frac{1}{{z}^{2}}}\left( {\eta}^{i}{\eta}_{i}\right)^{2}  +\Big(\textstyle{\frac{i}{ {z}^{2}}} {z}_{N} {\eta}_{i}{\rho^{MN}}_{\phantom{i}j}^{i} {\eta}^{j}\Big)^{2}\Big]}\\\nonumber
&&  \!\!\!\!\!\!\!
\sim\!\int \!\!D\phi D\phi^M\,e^{-  g\int dt ds\,[\textstyle\frac{1}{2}{\phi}^2+\frac{\sqrt{2}}{z}\phi\,\eta^2 +\frac{1}{2}({\phi}_M)^2-i\,\frac{\sqrt{2}}{z^2}\phi^M {z}_{N} \,\big(i \,{\eta}_{i}{\rho^{MN}}_{\phantom{i}j}^{i} {\eta}^{j}\big)]}~.
\end{eqnarray}
Above, in the second line we have  written  the Lagrangian for $\phi^M$ so to emphasize that it has an imaginary part,  due to the fact that the bilinear form in round brackets   is hermitian
\begin{equation}
\!\!\!\!\!\!
\Big(i\,\eta_i {\rho^{MN}}^i{}_j \eta^j\Big)^\dagger=-i(\eta^j)^\dagger({\rho^{MN}}^i{}_j)^*(\eta_i)^\dagger
=-i \eta_j\,{\rho^{MN}}_i{}^j\,\eta^i=i\eta_j\,{\rho^{MN}}^j{}_i\,\eta^i
\,,
\end{equation}
as follows from the properties of the $SO(6)$ generators. 
Since the auxiliary vector field $\phi^M$ has real support, the  Yukawa-term for it sets \emph{a priori} a phase problem, 
the  only  question being whether the latter is treatable via standard reweighing.  
After the transformation \eqref{HubbardStratonovich}, the corresponding Lagrangian reads  
\begin{eqnarray}\nonumber
\!\!\!\!\!\!\!\!\!\!\!\!\!\!\!
{\cal L} &=& \!\textstyle {| \partial_t {x} +\!\frac{m}{2}{x} |}^2 + \!\frac{1}{{ z}^4}{| \partial_s {x} -\!\frac{m}{2}{x} |}^2
+\! (\partial_t {z}^M + \!\frac{1}{2}{z}^M )^2 +\! \frac{1}{{ z}^4} (\partial_s {z}^M -\!\frac{m}{2}{z}^M)^2\\\label{Scuspquadratic}
&&+\!\frac{1}{2}{\phi}^2 +\!\frac{1}{2}({\phi}_M)^2+\!\psi^T O_F \psi\,
\label{final_continuum_L}
\end{eqnarray}
  with  $\psi\equiv({\theta}^i, { \theta}_i, {\eta}^i, {\eta}_i)$ and  
\begin{equation}
\nonumber
\!\!\!\!\!\!\!\!\!\!\!\!\!\!\!\!\!\!\!\!\!
\textstyle{
O_F =
\left(\!\!\!\! \begin{array}{cccc}
\!\!\!\! 0 &  \!\!\!\! i\partial_{t} & \!\!\!\! \textstyle{-\mathrm{i}\rho^{M}\left(\partial_{s}+\frac{m}{2}\right)\frac{{z}^{M}}{{z}^{3}}} & \!\!\!\! 0\\
\mathrm{i}\partial_{t} & 0 & 0 & -\mathrm{i}\rho_{M}^{\dagger}\left(\partial_{s}+\frac{m}{2}\right)\frac{{z}^{M}}{{z}^{3}}\\
\mathrm{i}\frac{{z}^{M}}{{z}^{3}}\rho^{M}\left(\partial_{s}-\frac{m}{2}\right) & 0 & 2\frac{{z}^{M}}{{z}^{4}}\rho^{M}\left(\partial_{s}{x}-m\frac{{x}}{2}\right) & i\partial_{t}-A^T\\
0 & \mathrm{i}\frac{{z}^{M}}{{z}^{3}}\rho_{M}^{\dagger}\left(\partial_{s}-\frac{m}{2}\right) &\mathrm{i}\partial_{t}+A & -2\frac{{z}^{M}}{{z}^{4}}\rho_{M}^{\dagger}\left(\partial_{s}{x}^*-m\frac{{x}}{2}^*\right)
\end{array}\!\!\!\! \right),}
\end{equation}
\begin{equation}
\label{Aoperator}
A  =\frac{1}{\sqrt{2}{z}^{2}}{\phi}_{M}\rho^{MN} {z}_{N}-\frac{1}{\sqrt{2}{z}}{\phi}\, +\mathrm{i}\,\frac{{z}_{N}}{{z}^{2}}\rho^{MN} \,\partial_{t}{z}^{M}~.
\end{equation}
%
The  quadratic fermionic contribution  resulting from linearization gives then formally a Pfaffian ${\rm Pf}\,O_F$, which - in order to enter the Boltzmann weight and thus  be interpreted as a probability  - should be  positive definite.
For this reason, one proceeds as follows
\begin{equation}\label{fermionsintegration}
\!\!\!\!
 \int \!\! D\Psi~ e^{-\textstyle\int dt ds \,\Psi^T O_F \Psi}={\rm Pf}\,O_F\equiv(\det O_F\,O^\dagger_F)^{\frac{1}{4}}= \int \!\!D\xi D\bar\xi\,e^{-\int dt ds\, \bar\xi(O_FO^\dagger_F)^{-\frac{1}{4}}\,\xi}~,
 \end{equation}
where the second equivalence obviously ignores potential phases or anomalies. 
%
%
  The values of the discretised (scalar) fields are assigned to each lattice site,  with periodic boundary conditions for all the fields except for antiperiodic temporal boundary conditions in the case of fermions. The discrete approximation of continuum derivatives are finite difference operators defined on the lattice. A Wilson-like lattice operator must be introduced, such that   fermion doublers are suppressed 
 and the one-loop constant $-3\ln 2/\pi$  in (\ref{cuspperturbative}) is recovered in lattice perturbation theory.   

\subsection{Simulations} 
\begin{figure}[t]
    \centering
           \includegraphics[scale=0.45]{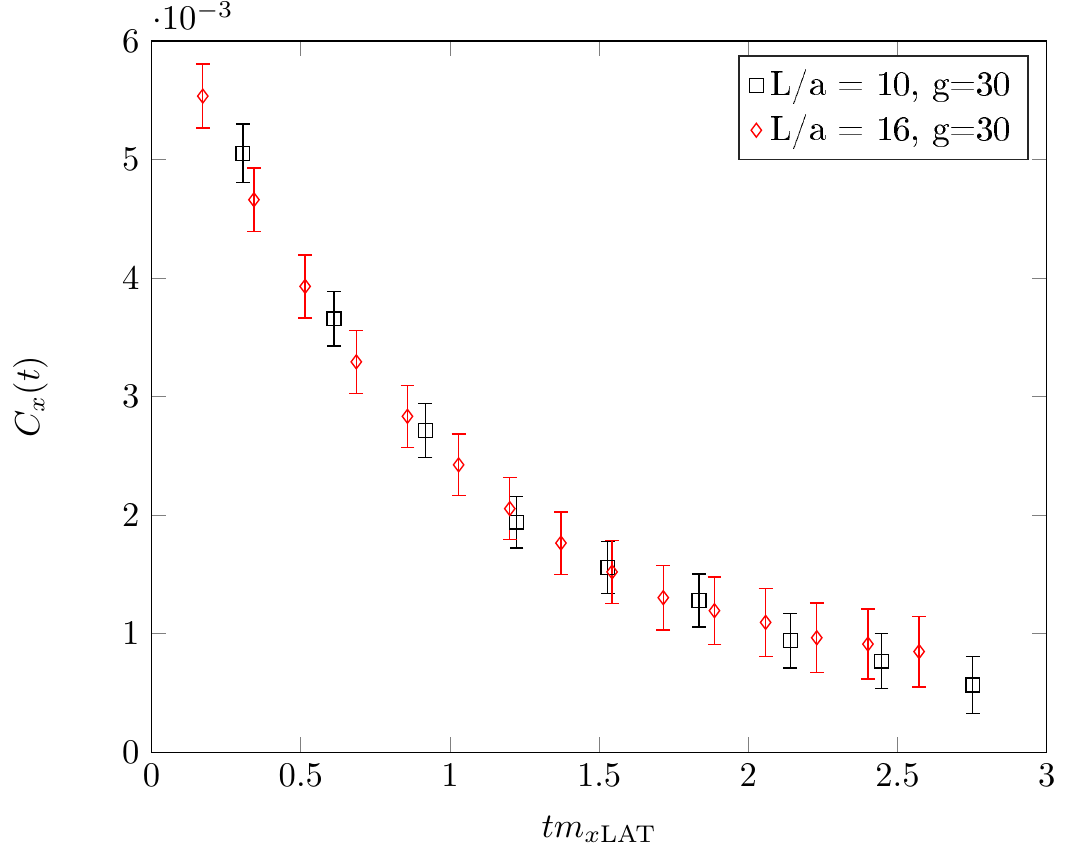}        
         \hspace{0.5cm}
         \hspace{0.5cm}
          \includegraphics[scale=0.45]{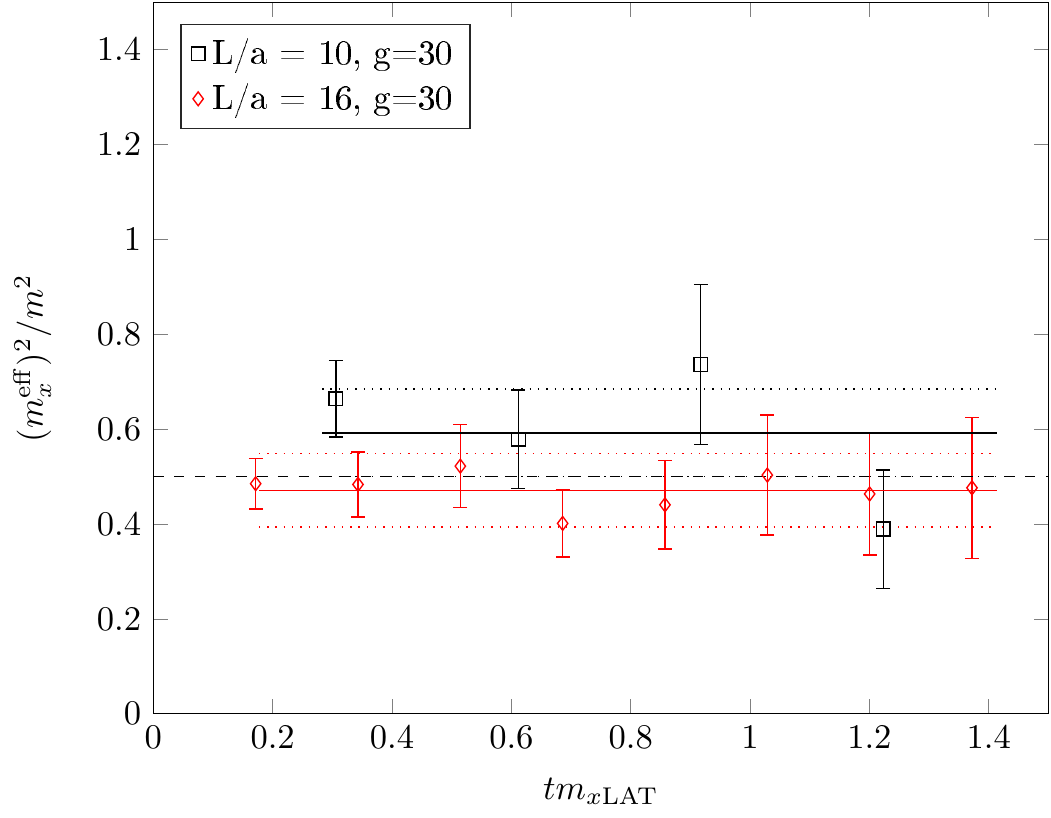}
           \hspace{0.5cm}
            \caption{Correlator $C_x(t)=\sum_{s_1,s_2} \langle x(t,s_1) x^*(0,s_2)\rangle$ of bosonic fields $x,x^*$ ({\bf left panel}) and corresponding effective mass $m^{\rm eff}_x=\frac{1}{a}\ln\frac{C_x(t)}{C_x(t+a)}$ normalized by $m^2$ ({\bf right panel}), plotted as functions of the time $t$ in units of ${m_x}_{\rm LAT}$ for different $g$ and lattice sizes. The flatness of the effective mass indicates that the ground state saturates the correlation function, and allows for a reliable extraction of the mass of the $x$-excitation.  
            Data points are masked by large errorbars for time scales greater than unity because the signal of the correlator degrades exponentially compared with the   statistical noise. } 
             \label{fig:correlatormass}
\end{figure}

The Monte Carlo evolution of the action \eqref{Scuspquadratic}   is generated by the standard Rational Hybrid Monte Carlo (RHMC) algorithm,  
with a rational approximation (Remez algorithm) for the inverse fractional power in the last equation of \eqref{fermionsintegration}, 
 as in \cite{McKeown:2013vpa}. 
 %
In the continuum model there are two parameters,  the  dimensionless coupling $g=\frac{\sqrt{\lambda}}{4\pi}$  and the mass scale $m$. 
In taking the continuum limit, the dimensionless physical quantities that it is natural to keep constant are the physical masses of the field excitations  rescaled by $L$, the spatial lattice extent. This is our line of constant physics.  For the example in~\eqref{cuspperturbative}, this means 
\begin{equation}\label{constantphysics}
L^2 \,m^2_{x}=\text{const}\,,\qquad~~\text{which leads to}\qquad~~L^2 \,m^2\equiv (N M)^2 =\text{const}\,,
\end{equation}
where we defined the dimensionless $M=m a$ with the lattice spacing $a$.  
The second equation in \eqref{constantphysics} relies first on the assumption that $g$ is \emph{not} renormalized, which is suggested by lattice perturbation theory. Second, one should investigate whether the second  relation in \eqref{cuspperturbative},  and the analogue ones for the other fields  of the model, are still true in the discretized model - \emph{i.e.} the physical masses undergo only a \emph{finite} renormalization. In this case, at each fixed $g$ fixing $ L^2~m^2$ constant would be enough to keep the rescaled physical masses constant, namely   no tuning of the ``bare'' parameter $m$ would be necessary.  
%
%
In~\cite{Bianchi:2016cyv}, we considered the example of bosonic $x,x^*$ correlators, 
whose asymptotic exponential decay is governed by the physical mass ${m_x}_{\rm LAT}$, as from the partially Fourier transformed
\be
\!\!\!\!\!\!\!
C_x(t;\,0)~ \stackrel{t\gg1}{\sim} ~e^{- t\, {m_x}_{\rm LAT}}, \qquad  {m_x}_{\rm LAT}=\lim_{T,\,t\to\infty}m^\textrm{eff}_{x}\equiv \lim_{T,\, t\to\infty, }\frac{1}{a} \log\frac{C_x(t;\,0)}{C_x(t+a;\,0)}.
\ee
On the lattice ${m_x}_{\rm LAT}$ is usefully obtained as a limit of an effective mass, the discretized logarithmich derivative above, 
that in Fig. \ref{fig:correlatormass} is measured as a function of the time t (in units of ${m_x}_{\rm LAT}$) for different lattice sizes. 
No ($1/a$) divergence is found, and in the  large $g$  region that we investigate  
the ratio considered approaches the expected continuum value $1/2$.   
%
%
%
%
Having this as hint corroborating the choice of the line of constant physics,  and because with the proposed discretization we recover in perturbation theory the one-loop cusp anomaly, we assume  that in the discretized model no further scale but the lattice spacing $a$  is present.  Any  observable $F_{\rm LAT}$ is therefore a function $F_{\rm LAT}=F_{\rm LAT}(g,N,M)$ of the  input (dimensionless) parameters $g=\frac{\sqrt{\lambda}}{4\pi}$, $N=\frac{L}{a}$ and  $M=a \,m$.
 At fixed coupling $g$ and fixed $m\, L\equiv M\,N$ (large enough so to keep finite volume effects $\sim e^{-m\,L}$  small),    $F_{\rm LAT}$ is evaluated for different values of $N$. The continuum limit  -- which we do not attempt here -- is then obtained extrapolating to infinite~$N$.

\begin{figure}[h]
    \centering
   \includegraphics[scale=1]{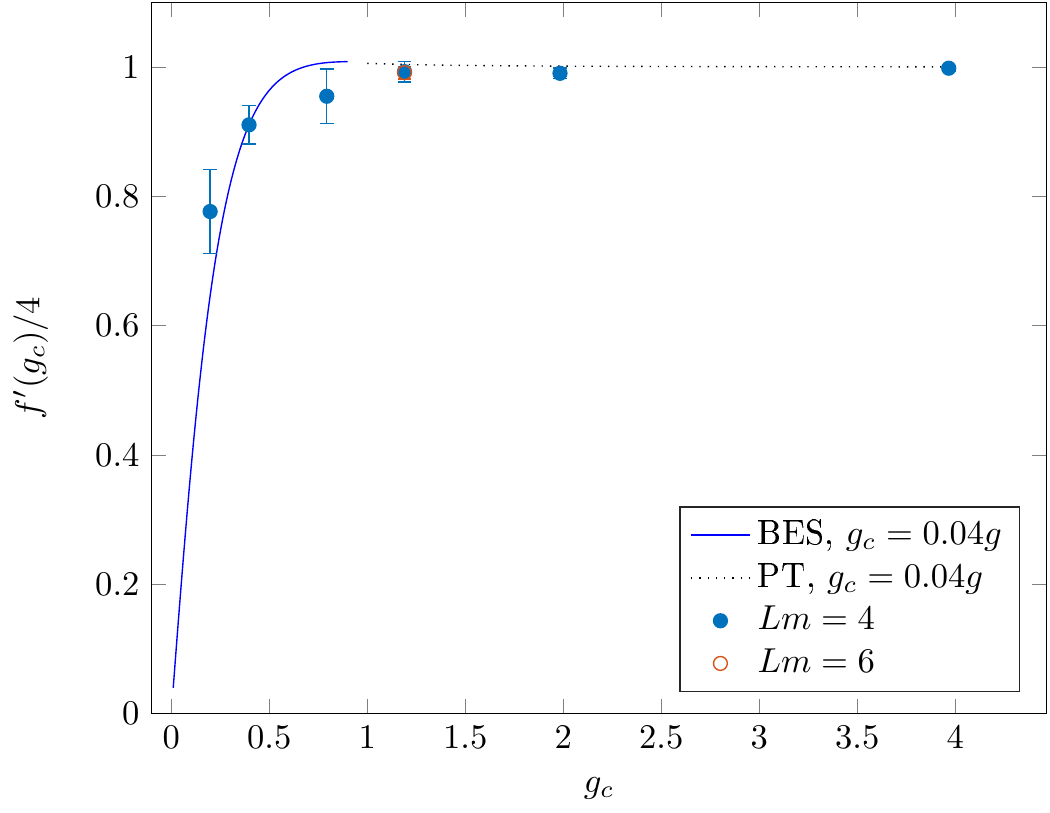}  
        \caption{Plot for $f'(g)/4$ as a function of the (bare) continuum coupling $g_c$ under the hypothesis that the latter is just a finite rescaling of the lattice bare coupling $g$ ($g_c=0.04\,g$).  The dashed line represents the first few terms in the perturbative series, the continuous line is obtained from a numerical solution of the BES equation and represents therefore the prediction from  the integrability of the model. The simulations  at $g=30$, $m\,L=6$ (orange point) are used for a check of the finite volume effects, that appear to be within statistical errors.} 
     \label{fig:action_fin_g}
\end{figure}

\begin{figure}[t]
   \centering
 \includegraphics[scale=0.55]{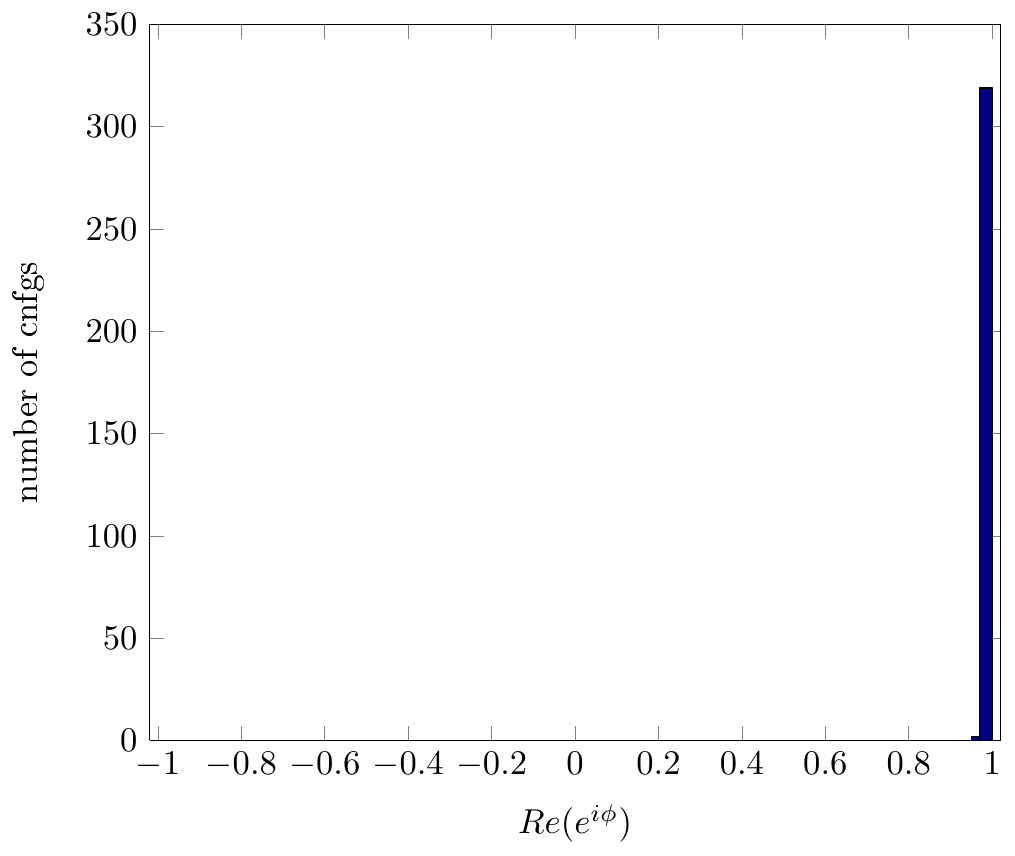}
  \includegraphics[scale=0.55]{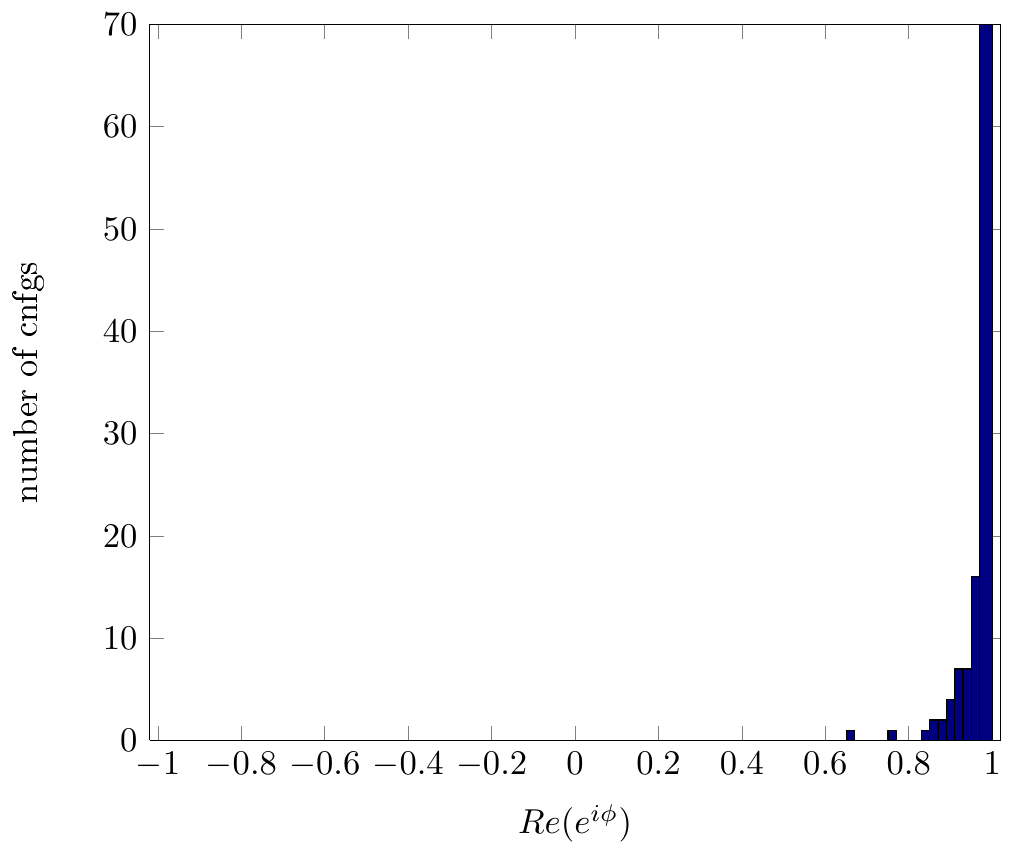}
   \includegraphics[scale=0.55]{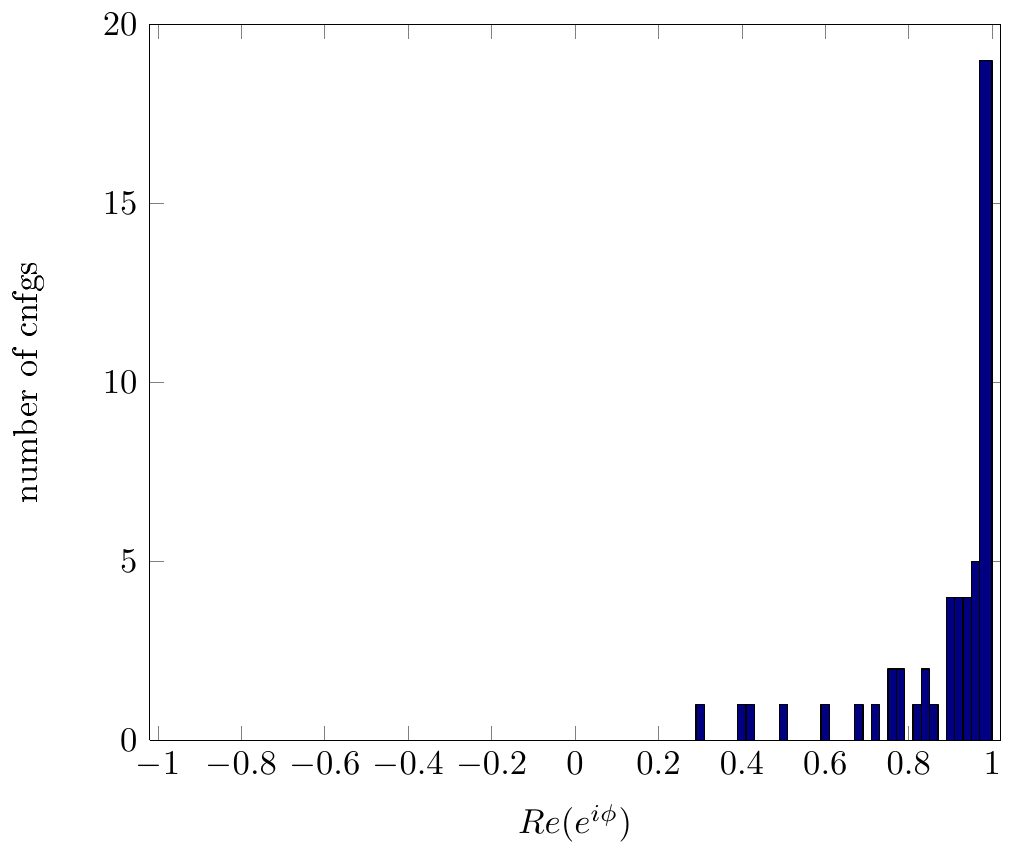}
    \includegraphics[scale=0.55]{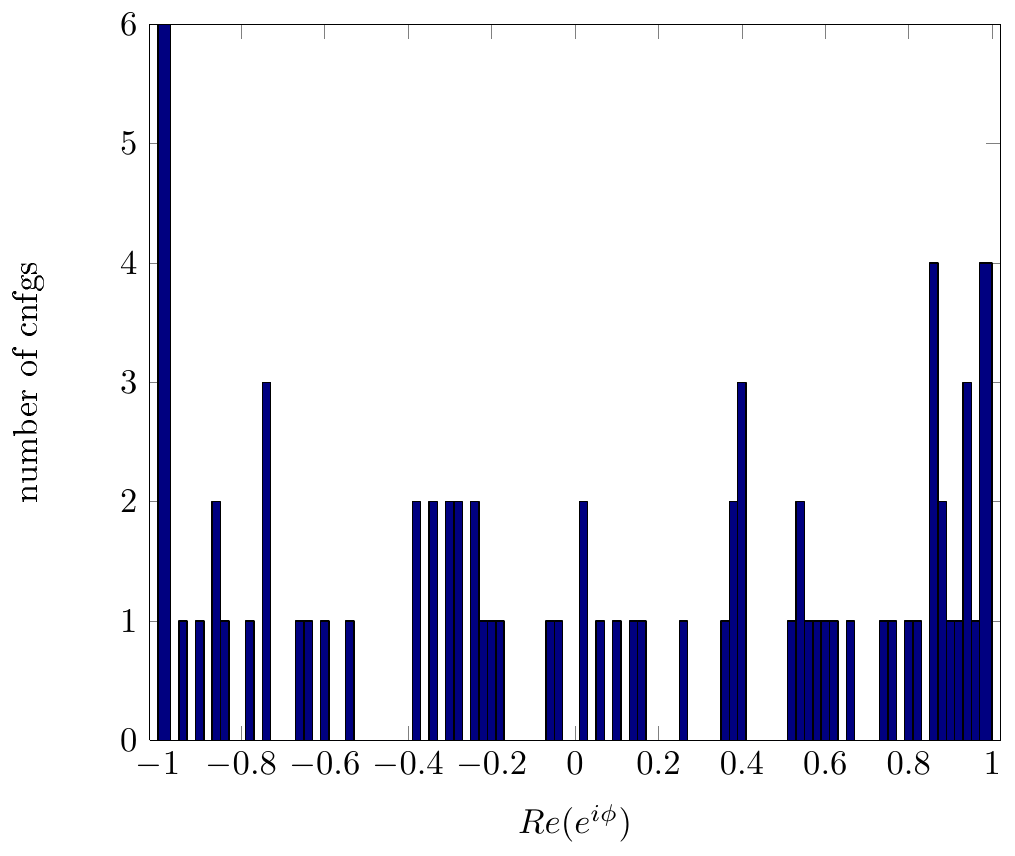}
 \caption{Histograms for the  frequency of the real part 
 of the   reweighting phase factor $e^{i\theta}$ of the Pfaffian ${\rm Pf}\,O_F=|(\det O_F)^{\frac{1}{2}}|\,e^{i\theta}$, based on the ensembles generated  at  $g=30,10,5,1$ (from left to right, top to down) for $L/a=8$. }
\label{fig:phase}
\end{figure}

In measuring  the action \eqref{vevaction} on the lattice, we are supposed to recover the following general behavior 
\begin{equation} \label{fit}
\frac{\langle S_{\rm LAT}\rangle}{N^2} = \frac{c}{2}+\frac{1}{8}\,M^2\,g \,f'(g)\,, 
\end{equation}
 where we have reinserted the parameter $m$, used  that $V_2=a^2\,N^2$ and added a constant  contribution  in $g$ which takes into account possible coupling-dependent Jacobians relating the (derivative of the) partition function on the lattice to the one in the continuum. 
Measurements for the ratio
$ 
\frac{\langle S_{LAT}\rangle-c\, N^2/2}{M^2\,N^2\,g/2} = \frac{f'(g)}{4} ~
$
 are, at large $g$,  in good agreement with $\frac{c}{2}=7.5(1)$, consistently with the with the counting of those degrees of freedom which appear quadratically, and multiplying $g$, in the action -- the number of bosons~\footnote{In lattice codes, it is conventional to omit the coupling form the (pseudo)fermionic part of the action, since this is quadratic in the fields and hence its contribution in g can be evaluated by a simple scaling argument. }.
 Having determined with good precision the coefficient of the divergence, one proceeds first fixing it to be exactly $c = 15$ and subtracting it from the action. At large $g$,  
 a good agreement is found with the leading order prediction in  \eqref{cuspperturbative} for which $ f'(g) =4$.    For lower values of $g$ 
 one observes deviations 
 that obstruct the continuum limit and signal  the presence of further quadratic ($\sim N^2$) divergences.  It seems natural  to relate these power-divergences to those arising in continuum perturbation theory and mentioned in the previous Section, where they are                                                            usually set to zero using dimensional regularization. From the perspective of a hard cut-off regularization like the lattice one, this is related to the emergence in the continuum limit of power divergences -- quadratic, in the present two-dimensional case -- induced by mixing of the (scalar) Lagrangian with the identity operator under UV renormalization.    %
One may proceed  with a non-perturbative subtraction of these divergences. 
 A simple look at  Fig. \ref{fig:action_fin_g} shows that, in the perturbative region, our analysis -- and the related assumption for the finite rescaling of the coupling -- is in good qualitative agreement with the integrability prediction. 
About direct comparison with the perturbative series, the plot in Fig. \ref{fig:action_fin_g} does not catch the minimal upward trend of the first correction to the expected large g behavior $f'(g)/4 \sim1$ - we are considering the derivative of eq. (21) and the first correction is (positive and) too small, about 2 percent, if compared to the statistical error. 
Notice that, again under the assumption that such simple relation between the couplings exists  -- something that within our error bars cannot be excluded  --  the nonperturbative regime beginning with $g_c=1$ would start at $g=25$, implying that our simulations at $g=10, 5$  would  already test a fully non-perturbative regime of the string sigma-model under investigation.   
%
%
%
In proximity to $g\sim 1$,  
severe  fluctuations appear in the averaged \emph{complex} phase of the Pfaffian    -- see Figure \ref{fig:phase} -- signaling the sign problem mentioned above. 
Interestingly, at least some steps in the direction of solving this problem can be done analytically~\cite{Forini:2017whz, toappear_lattice}. A new auxiliary field representation of the four-fermi term may be realized, following an algebraic manipulation from which an  hermitian Lagrangian linearized in fermions results, leads to a Pfaffian of the quadratic fermionic operator $O_F$ which is \emph{real},  $({\rm Pf}\,O_F)^2=\det O_F \geq0$. Although a sign ambiguity remains, as the Pfaffian is still not positive definite, ${\rm Pf} \,O_F=\pm \det O_F$, this is an important advancement in the efficiency of the simulations, as it allows eliminating  systematic errors and identifying with precision the region of parameter space where information on nonperturbative physics may be captured. 

 \section{Discussion and outlook}
\label{sec:conclusions}

We have reviewed and discussed perturbative and non-perturbative approaches to the quantization of the Green-Schwarz string in AdS backgrounds with RR-fluxes,  with an emphasis on the use of direct quantum field theory methods and on the cross-fertilization of theoretical tools well-established  in gauge field theories to the string worldsheet context.

In dealing with sigma-model perturbation theory, crucial subtleties appear in evaluating regularised functional determinants for string fluctuations and the computational technology for them has to be carefully adjusted to the problem at hand. 
It would be important to develop a diffeomorphism-preserving regularization scheme which retains the efficiency of the Gelf'and-Yaglom method for one-dimensional cases, and  
extend such techniques to regularised super-traces and super-determinants so to address a uniform way of treating BPS and non-BPS observables~\footnote{See however the recent developments in~\cite{Cagnazzo:2017sny}.}.
It would be extremely interesting to elucidate the role of the measure (structure and normalization) in the string path integral for supersymmetric configurations. 

Going beyond perturbation theory, a new research line  has been addressed, which employs  Montecarlo simulations to investigate observables defined on suitably discretized euclidean string worldsheets. At a fundamental level,  this is the natural setup for verifying with unequaled definiteness the holographic conjecture \emph{and} the exact methods that ``solve'' various sectors of the AdS/CFT system.  Lattice methods are also  the most suitable candidates for the study of several observables and backgrounds for which alternative techniques to go beyond perturbation theory are not existing (string backgrounds which are not classically integrable) or yet at a preliminary stage (correlators of string vertex operators and dual gauge theory correlation functions). 

It is important to emphasize that the analysis here carried out is far from being a non-perturbative definition, \`a la Wilson lattice-QCD, of the Green-Schwarz  worldsheet string model. For this purpose one should work with a Lagrangian which is invariant under the local symmetries - bosonic diffeomorphisms and $\kappa$-symmetry -  of the model,  while as mentioned we make use of an  action which fixes them all. There is however a number of reasons which make this model  interesting for lattice investigations, within and hopefully beyond the community interested in holographic models.  

 As computational playground this is an  interesting one on its own, allowing in principle for explicit investigations/improvements of algorithms: a highly-nontrivial two-dimensional model with four-fermion interactions, for which  relevant observables have not only, through AdS/CFT,  an explicit analytic  strong coupling expansion -- the perturbative series in the dual gauge theory  -- but also, through AdS/CFT \emph{and} the assumption of integrability, an explicit numerical prediction at all couplings.

The results discussed here open the way to a variety of further explorations and developements. 
A natural evolution consists in treating strings propagating in those backgrounds (the ten-dimensional $AdS_4\times CP^3$, $AdS^3\times S^3\times T^4$, $AdS_3\times S^3\times S^3 \times S^1$ supported by RR fluxes) relevant for lower-dimensional formulations of the correspondence, for which several predictions exist from integrability, and for which an independent  their AdS-light-cone gauge-fixed Lagrangians are expected to be considerably more involved than in the prototypical case, but still with vertices at most quartic in fermions. 
 In all the novel cases of study  the presence of massless fermionic modes is expected to require an ad-hoc treatment, one possibility being to work in a finite-volume setting like the Schr\"odinger functional scheme. It would be crucial a thorough study of the possible sign problem - related to the absence of positive definiteness of the fermion Pfaffian - that is likely to appear at large values of the string tension as in the prototypical case. This may consists in carving out the region of parameter space where the sign ambiguity is not severe and clarifying whether non-perturbative physics is obtainable. One may then verify the possibility of track down this ambiguity to the behaviour of a smaller set of degrees of freedom -- such analysis may profit, at least pedagogically, from recent  progress on the analysis of the sign problem in (considerably simpler) models with quartic fermionic interactions~\cite{Catterall:2015zua, Alexandru:2016ejd}. 
Also, it would be very interesting to explore the discretization of the gauge-fixed string action of~\cite{Kallosh:1998ji},  whose relevance from the point of view of the string/gauge gravity correspondence is far less clear, but that being only quadratic in fermions may lead to considerable simplifications in the general analysis. 

\section*{Acknowledgements}

It is a pleasure to thank Lorenzo Bianchi, Marco S. Bianchi, Alexis Br\'es, Ben Hoare, Valentina Giangreco M. Puletti, Luca Griguolo, Bjoern Leder, Michael Pawellek,  Domenico Seminara, Philipp Toepfer, Arkady A. Tseytlin and Edoardo Vescovi for the very nice collaboration on Refs.~\cite{Bianchi:2013nra, Bianchi:2014ada, Forini:2014kza, Forini:2015mca, Forini:2015bgo,  Bianchi:2016cyv, Forini:2016sot, Forini:2017mpu, Forini:2017whz, toappear_lattice}, on which this review is based. In particular, I am grateful to the long-period members of the Emmy Noether Group ``Gauge Fields from Strings'' -- Ben Hoare, Lorenzo Bianchi and Edoardo Vescovi -- for joining me and my research program, and making possible to achieve several relevant goals in our given time frame. This research was largely supported by the German Research Foundation (DFG) through the Emmy Noether Group 31408816. The novel, interdisciplinary program of discretization and simulation of the Green-Schwarz superstring would have not been possible without the support of the Collaborative Research Centre ``Space-time-matter'' SFB 647 -- subproject C5 ``AdS/CFT Correspondence: Integrable Structures and new Observables'' -  involving several scientists at Berlin and Postdam Universities.

\bibliographystyle{nb}

\bibliography{Ref_SFB_Essay}
\end{document}